\documentclass[apj,referee]{emulateapj}
\usepackage{amssymb}
\usepackage{graphicx,graphics}			
\usepackage{epsfig}						
\usepackage{natbib}
\usepackage{times}

\newcommand{\RomanNumeralCaps}[1]
    {\MakeUppercase{\romannumeral #1}}

\usepackage{pslatex}
\usepackage{graphicx}
\usepackage{gensymb}

\shorttitle{Cluster environments of AGN}
\shortauthors{Kar Chowdhury et al.}

\begin{document}

\title[Cluster environments of AGN]{Cosmological Simulation of Galaxy Groups and Clusters-I: Global Effect of Feedback from Active Galactic Nuclei} 

\author{Rudrani Kar Chowdhury\altaffilmark{1}, Suchetana Chatterjee\altaffilmark{1}, 
Anto. I. Lonappan\altaffilmark{2,1,3}, Nishikanta Khandai\altaffilmark{4} and
Tiziana DiMatteo \altaffilmark{5}
}
\altaffiltext{1}{Department of Physics, Presidency University, Kolkata, 700073, India}
\altaffiltext{2}{Astrophysics and Cosmology, Scuola Internazionale Superiore di Studi Avanzati, via Bonomea, 265 - 34136 Trieste, Italy}
\altaffiltext{3}{Institute for Fundamental Physics of the Universe (IFPU), Via Beirut 2, 34014 Trieste, Italy}
\altaffiltext{4}{School of Physical Sciences, National Institute of Science Education and Research, HBNI, Jatni - 752050, India }
\altaffiltext{5}{McWilliams Center for Cosmology, Carnegie Mellon University, Pittsburgh, PA 15213 USA }

\begin{abstract}
In this study we quantify the properties of the gas and dark matter around active galactic nuclei (AGN) in simulated galaxy groups and clusters and analyze the effect of AGN feedback on the surrounding intra-cluster (group) medium. Our results suggest downsizing of AGN luminosity with host halo mass, supporting the results obtained from clustering studies of AGN. By examining the temperature and density distribution of the gas in the vicinity of AGN we show that due to feedback from the central engine, the gas gets displaced from the centre of the group/cluster resulting in a reduction of the density but an enhancement of temperature. We show that these effects are pronounced at both high and low redshifts and propose new observables to study the effect of feedback in higher redshift galaxies. We also show that the average stellar mass is decreased in halos in the presence of AGN feedback confirming claims from previous studies. Our work for the first time uses a fully cosmological-hydrodynamic simulation to evaluate the global effects of AGN feedback on their host dark matter halos as well as galaxies at scales of galaxy groups and clusters.

\end{abstract}

\section{Introduction}
It is currently believed that every massive galaxy in the Universe harbors a central supermassive black hole (SMBH) of mass ranging between $10^{6}-10^{9} M_{\odot}$ \citep[e.g.,][]{soltan82,m&f01,Ferrareseetal2000ApJ,gultekinetal09}. Some of them have active accretion discs and they grow by accreting gas from their surroundings. These classes of SMBH are known as active galactic nuclei (AGN). As they grow by accreting matter from their neighbouring environments they release a large amount of energy into their surroundings. Some fraction of the radiated energy couples to the surrounding gas in a process known as AGN feedback \citep[e.g.,][]{SilkRees1998A&A, CiottiOstriker2001ApJ, NathRoychowdhury2002MNRAS, Kaiseretal2003MNRAS, Nulsenetal2004cosp, Dimatteoetal2005Natur, Springeletal2005MNRAS, Coxetal2006ApJ, Raychaudhuryetal2009ASPC, Chaudhurietal2013ApJ, Chatterjeeetal2015PASP, Costaetal2015MNRAS, Pennyetal2018MNRAS, Harrisonetal2018NatAs}.

Many studies show that AGN feedback has observable effects on galaxy formation known as AGN-galaxy co-evolution in the literature \citep[e.g.,][]{KauffmannHaehnelt2000MNRAS, WyitheLoeb2003ApJ, Marconietal2004MNRAS, Shankaretal2004MNRAS, Cattaneoetal2006MNRAS, Crotonetal2006MNRAS, Hopkinsetal2006ApJ, Lapietal2006ApJ, Dimatteoetal2008ApJ, BoothSchaye2009MNRAS, Volonterietal2011ApJ, ConroyWhite2013ApJ, Lapietal2014ApJ, Caplaretal2015ApJ, Oogietal2016MNRAS, Lanzuisietal2017A&A, Mutluetal2018MNRAS}. Several properties of the host galaxy or the surrounding environment of the central engine is linked with the central SMBH itself. For example, there exists strong correlation between the central black hole mass and the stellar velocity dispersion \citep[e.g.,][]{Ferrareseetal2000ApJ, Tremaineetal2002ApJ, FerrareseFord2005SSRv}. Also other properties of the galaxy such as bulge luminosity \citep[e.g.,][]{Dressler1989IAUS, KormendyRichstone1995ARA&A, MarconiHunt2003ApJ, Graham2007MNRAS}, bulge mass \citep[e.g.,][]{Magorrianetal1998AJ, HaringRix2004ApJ}, Sersic index \citep[e.g.,][]{GrahamDriver2007ApJ}, kinetic energy of the random motion of the bulge \citep[e.g.,][]{FeoliMancini2009ApJ}, gravitational binding energy \citep[e.g.,][]{AllerRichstone2007ApJ}, virial mass of the galaxy \citep[e.g.,][]{Ferrareseetal2006ApJ} are tightly coupled to the mass of the central SMBH. Studies show that the effective radius of galaxies is correlated with the residual of $\rm M_{BH}-\sigma$ and $\rm M_{BH}-L_{Bulge}$ relations \citep[e.g.,][]{MarconiHunt2003ApJ, Hopkins2007ApJ, Hopkinsb2007ApJ}. It has been also suggested that there exists a plane of correlation between black hole mass, galaxy size and bulge mass, known as the black hole fundamental plane \citep[e.g.,][]{Hopkinsb2007ApJ, Beifiorietal2012MNRAS, Saikiaetal2015MNRAS, Vandenbosch2016ApJ}.

Other observable effects of AGN feedback include the $\rm L_{X}-T$ relation in galaxy groups and clusters \citep[e.g.,][]{ArnaudEvrard1999MNRAS, NathRoychowdhury2002MNRAS, ScannapiecoOh2004ApJ, Lapietal2005ApJ, PetersonFabian2006PhR, Thackeretal2009ApJ, Puchweinetal2010MNRAS, Bharadwajetal2015A&A, Voitetal2018ApJ}, Sunyaev-Zeldovich \citep[SZ:][]{Sunyaev1972CoASP} effect \citep[e.g.,][]{Bhattacharyaetal2008MNRAS, Chatterjeeetal2008MNRAS, Chatterjeeetal2010ApJ, Ruanetal2015ApJ, Verdieretal2016A&A, Crichtonetal2016MNRAS, Spaceketal2016ApJ, lacyetal19}, SZ power spectrum \citep{ChatterjeeKosowsky2007ApJ, Scannapiecoetal2008ApJ, Battagliaetal2010ApJ}, reduced star formation rate \citep[e.g.,][]{Kauffmannetal2003MNRAS, Vitaleetal2013A&A, Vaddietal2016ApJ, Terrazasetal2017ApJ, Kakkadetal2017mnras} which have been addressed in the literature.

\begin{figure*}
 \begin{tabular}{c}
     \includegraphics[width=8.9cm]{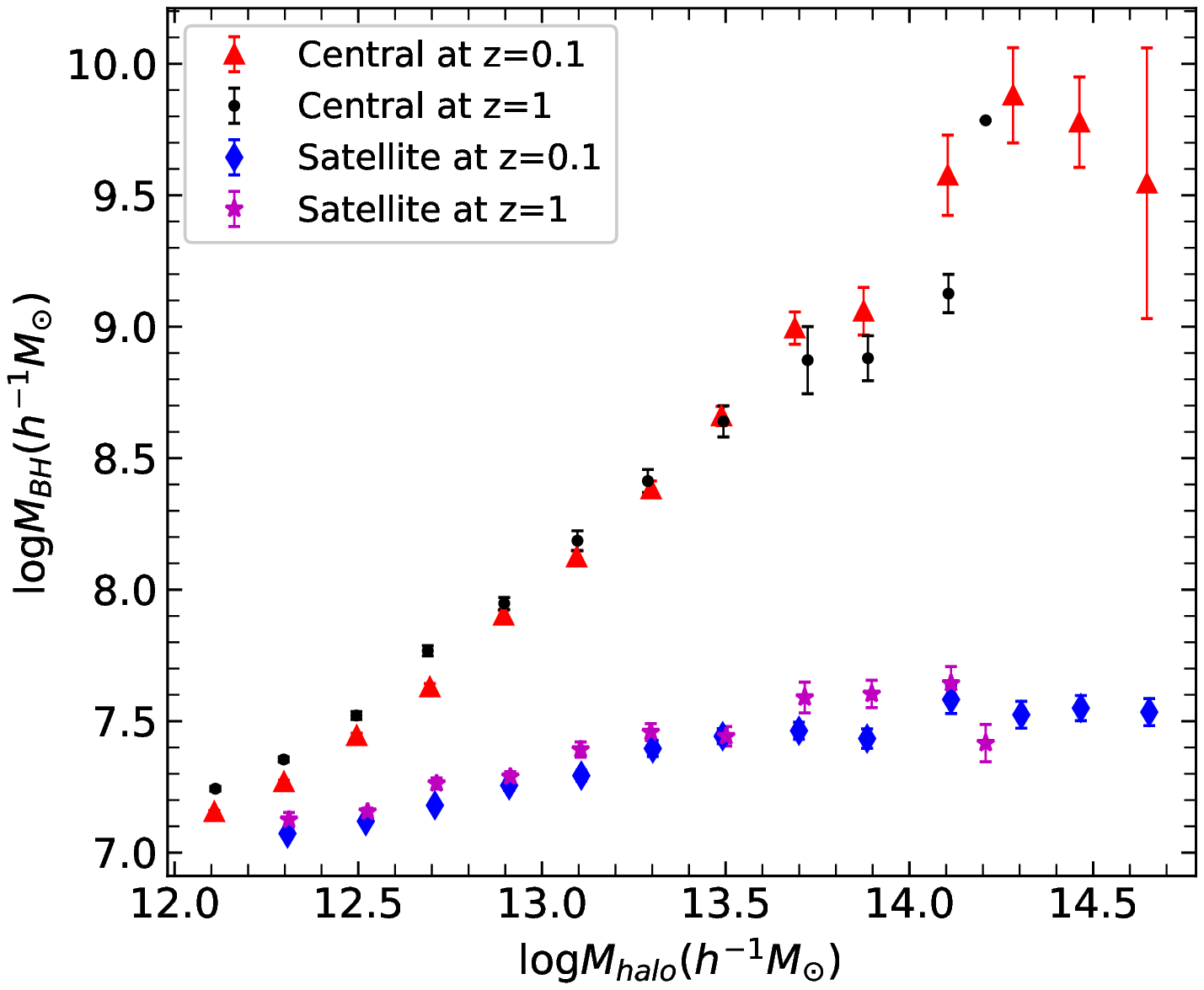}
      \includegraphics[width=8.9cm]{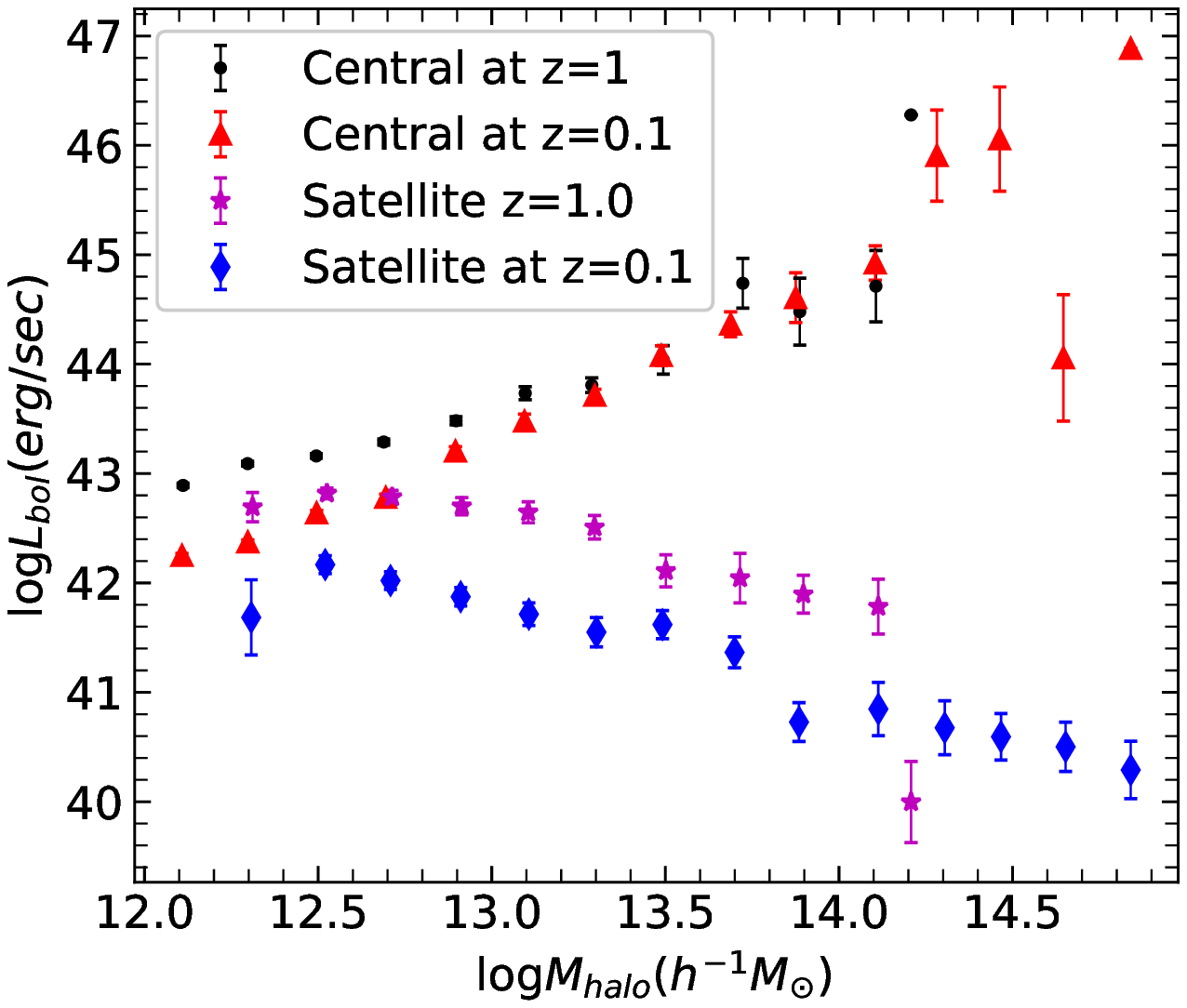}
     \end{tabular}
 \caption{Left panel : $\rm M_{halo}-M_{BH}$ relation in the K15 simulation at $z=1$ and at $z=0.1$ . Black dots and magenta stars show the central and satellite black holes respectively at $z=1$, while red triangles and blue diamonds represent the central and satellite black holes respectively at $z=0.1$. We do not see any redshift dependance of these correlations. Central BH mass increases steeply with their host halo mass while the satellite BH mass increases very slightly with the halo mass and then gets saturated at the higher host halo mass range. Right panel : $\rm M_{halo}-L_{bol}$ relation in the K15 simulation at $z=1$ and $z=0.1$. Color scheme is same as the left panel. Clearly the luminosities of the satellite black holes decrease in the higher halo mass bins for both redshifts. We propose that this downsizing effect is due to AGN feedback. See \S 3 and \S 4 for more discussions.}
  \label{fig:scaling2}
 \end{figure*}

Theoretical models of this paradigm of co-evolution has been explored by numerous groups  \citep [e.g.,][]{Granatoetal2004ApJ, Vittorinietal2005MNRAS, Hopkinsetal2006ApJS, Monacoetal2007MNRAS, Pelupessyetal2007ApJ, Johanssonetal2009ApJ, Aversaetal2015ApJ, LuMo2015ApJ, Yuanetal2015ApJ, Biernackietal2017MNRAS}. All of these models explain the observed correlation in terms of strong feedback from the AGN \citep{SilkRees1998A&A, Dimatteoetal2005Natur, Coxetal2006ApJ, McCarthyetal2010MNRAS, HeckmanBest2014ARA&A, Choietal2015MNRAS, Steinbornetal2015MNRAS, Baronetal2017MNRAS, Yangetal2018MNRAS}. Theoretical studies have been also undertaken to explain the effect of AGN feedback on the surrounding hot gas in galaxy groups and clusters \citep[e.g.,][]{Hambricketal2011ApJ, Gasparietal2012MNRAS, Choietal2015MNRAS}. To study the effect of AGN feedback on large scale structure and to evaluate its importance on the evolutionary history of the Universe, AGN feedback has been introduced in cosmological simulations \citep[e.g.,][]{Zannietal2005A&A, Sijackietal2007MNRAS, Pelupessyetal2007ApJ, Dimatteoetal2008ApJ, Vogelsbergeretal2014Natur, Vogelsberger2014MNRAS, Sijackietal2015MNRAS, Geneletal2014MNRAS, Chatterjeeetal2008MNRAS, Bhattacharyaetal2008MNRAS, Johanssonetal2009ApJ, McCarthyetal2010MNRAS, Hirschmannetal2014MNRAS, Nelsonetal2015MNRAS, Steinbornetal2015MNRAS, Yuanetal2015ApJ, Schayeetal2015MNRAS, Liuetal2016MNRAS, Biernackietal2017MNRAS, Beckmannetal2017MNRAS, Scholtzetal2018MNRAS, Peiranietal2019MNRAS}.
 
\begin{table*}
\begin{center}
\begin{tabular}{| p{2cm}| p{2 cm}| p{1.5 cm} | p{2.5 cm}| p{2 cm}| p{2 cm}| p{2 cm}|}
\hline
Simulation & Boxsize ($h^{-1}Mpc$) & {$N_{p}$} & $m_{DM} (h^{-1}{M_{\odot}})$ & $m_{gas} (h^{-1}{M_{\odot}})$ & $\varepsilon (h^{-1}kpc)$ & $z_{end}$\\
\hline
D08 & $33.75$ & $2\times{216^3}$ & $2.75\times{10^8}$ & $4.24\times{10^7}$ & $6.25$ & $1.0$ \\
\hline
K15 & $100$ & $2\times{1792^3}$ & $1.1\times{10^7}$ & $2.2\times{10^6}$ & $1.85$ & $0.0$\\
\hline
\end{tabular}
\caption{The simulation parameters: $ N_{p}$, $ m_{DM}$, $ m_{gas}$, $\varepsilon$, $z_{end}$ represent the total number of particles (dark matter+gas), mass of dark matter particles, initial mass of gas particles, gravitational softening length and the final redshift respectively.}\label{table2}
\end{center}
\end{table*}

Recently \citet{Khandaietal2015MNRAS} ran a cosmological simulation MassiveBlack- \RomanNumeralCaps{2} (MB\RomanNumeralCaps{2}) to study a large representative volume of the Universe with a spatial resolution of $\sim few$ kpc. Using this simulation we study the effect of AGN feedback on its surrounding medium by evaluating the correlation between AGN activity with the properties of the intra-cluster medium (ICM) as well as the host dark matter halos of AGN.
The paper is structured as follows. In ${\$2}$ we briefly discuss the simulation and the data used for our work. In ${\$3}$ , we present the results. In ${\$4}$, we discuss and summarize our results. 

\begin{figure*}
\begin{center}
\begin{tabular}{c}
\includegraphics[width=16.0cm]{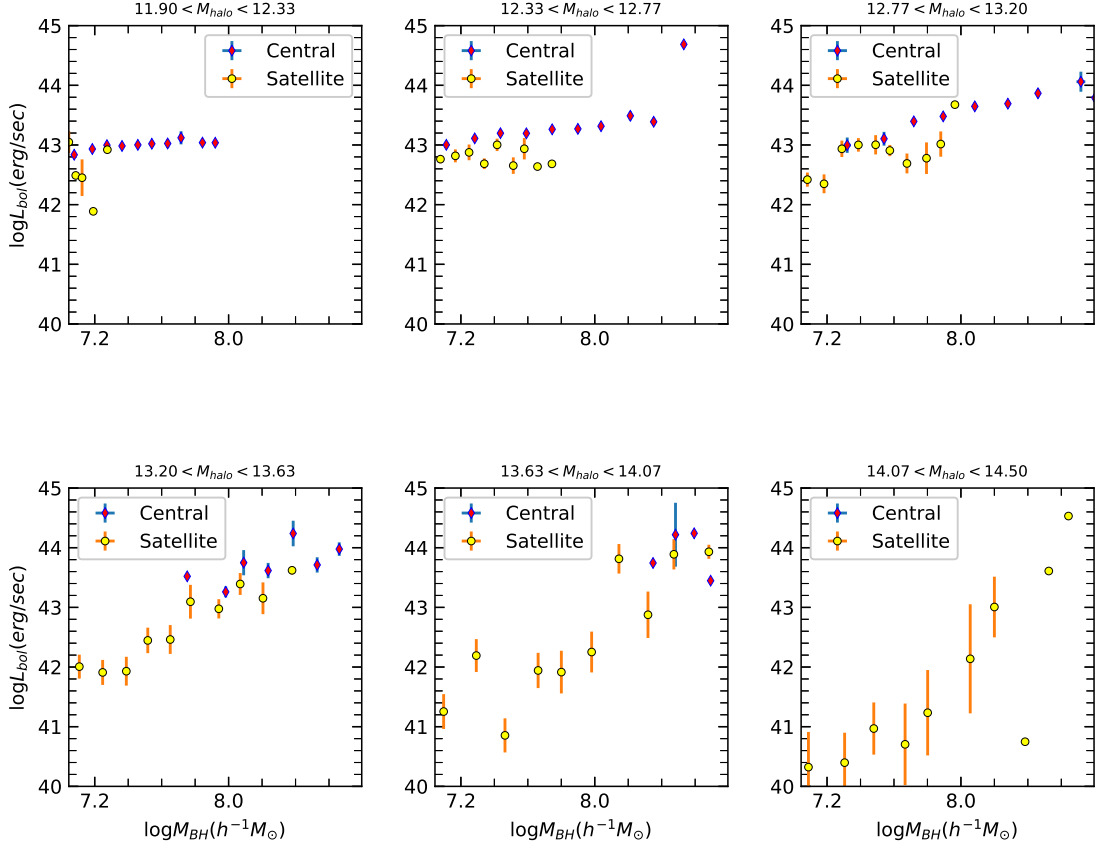}
\end{tabular}
\caption{Conditional probability distribution of the $\rm M_{BH}-L_{bol}$ correlations in the simulation at $z=1$. Blue diamonds and yellow circles show the central and satellite black holes repectively. Central AGN luminosity remains fairly constant with $\rm M_{BH}$ in the first two halo mass bins, then increasing slightly in the higher halo mass scales. However, upto a mass scale of $\approx 10^{13.2}h^{-1}$M$_{\odot}$, luminosity of the satellite AGN remains fairly constant. Above this mass scale there is a spread in the distribution of satellite AGN luminosities while the luminosities are typically low at the highest mass bin. The results are discussed in \S 3 and \S 4}
\label{fig:combine}
\end{center}
 \end{figure*}

\section{Simulation}
The simulation for this work uses an extended version of the parallel cosmological Tree Particle Mesh-Smoothed Particle Hydrodynamics code GADGET2 \citep{Springel2005MNRAS} and an upgraded version of it (GADGET3). Both simulations are based on the Lambda cold dark matter (${\Lambda}CDM$) cosmology with cosmological parameters adopted from \citet{Spergeletal2003ApJS} and \citet{Komatsuetal2011ApJS} respectively. We note that the different cosmological parameters will not affect the results of this paper. We refer the reader to \cite{Lietal2007ApJ} and \cite{Sijackietal2007MNRAS} for a detailed discussion in this regard. In this work we use two versions of the simulations namely: by \citealt{Dimatteoetal2008ApJ} (D08 hereafter) and 
\citealt{Khandaietal2015MNRAS} (K15 hereafter). D08 uses a simulation box of size 33.75 Mpc while the box size is much bigger for K15 (100 Mpc).

Both simulations have dark matter and gas dynamics. The simulations also include radiative gas cooling, star formation, black hole growth and feedback. The gas dynamics is modeled using the Lagrangian smoothed particle hydrodynamics (SPH) technique \citep{Monaghan1992ARA&A} and radiative cooling and heating processes are modeled using the prescription of \citet{Katzetal1996ApJS}. As these are cosmological volume simulations, the spatial resolution limits us in probing physical scales of star formation or black hole accretion. Hence to model these processes approximation schemes have been introduced. Star formation and supernova feedback are implemented in the simulation through a sub resolution multiphase model developed by \citet{SpringelHernquist2003MNRAS}.

Black hole accretion and feedback is modeled according to the prescription of \citet{Dimatteoetal2008ApJ} and \citet{Dimatteoetal2005Natur}. Black holes are assumed as collisionless sink particles that can grow by accreting matter from the intervening medium or through galaxy mergers. The Bondi-Hoyle spherical accretion relation \citep{BondiHoyle1944MNRAS, Bondi1952MNRAS} is used to quantify the accretion rate of the black hole. The accretion rate of gas on to the black hole is given by

\begin{equation}
\rm ~~~~~~~~~~~~~~~~~~~~~~~~~~~~~~~~\dot{M}_{BH}=4\pi\frac{(G^{2}{M_{BH}}^2\rho)}{({{{c_{s}}^2}+{v^2})}^{\frac{3}{2}}},
\label{eq:bondi}
\end{equation}
where $\rho$ and $c_{s}$ are the density and speed of sound of the surrounding gas respectively, $v$ is the velocity of the black hole relative to the surrounding gas and $\rm G$ is the universal gravitational constant. We note that due to limited resolution, the physical parameters in the Bondi-Hoyle relation are adopted from larger scales and adjusted with an appropriate boost factor \citep{Dimatteoetal2005Natur, Springeletal2005MNRAS, Pelupessyetal2007ApJ, Dimatteoetal2008ApJ}.

\begin{figure*}
\begin{center}
\begin{tabular}{c}
\includegraphics[width=16.0cm]{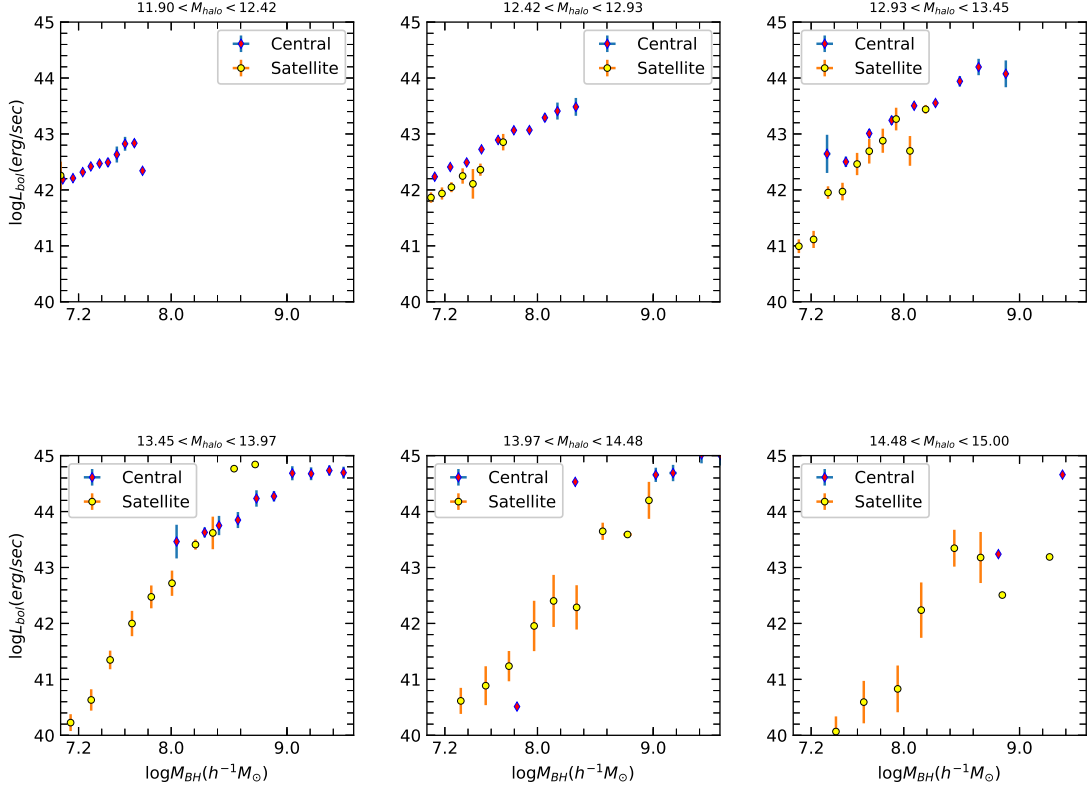}
\end{tabular}
\caption{Conditional distribution of the $\rm M_{BH}-L_{bol}$ correlations in the simulation at $z=0.1$. Blue diamonds and yellow circles represent the central and satellite AGN respectively. Luminosity of the central AGN slighltly increases with $\rm M_{BH}$, while there is a wide spread in the distribution of satellite AGN luminosities with the very low luminosity AGN in the high mass bins. The results are discussed in \S 3 and \S 4}
\label{fig:combine0.1}
\end{center}
 \end{figure*} 
 
The bolometric radiation coming out of the black hole is given by $\rm L_{bol}=\eta\dot{M}_{BH}c^2$, where $\eta$ is the canonical efficiency and its value is taken to be 0.1 for a radiatively efficient thin disc accretion \citep{ShakuraSunyaev1973A&A}. A part of this radiated energy gets associated with the local gas and is deposited on every gas particle according to a kernel function, as feedback energy $\rm E_{f}$ such that $\rm \dot{E}_{f}=\varepsilon_{f}L_{bol}$ \citep{Dimatteoetal2008ApJ}. $\rm \varepsilon_{f}$ is the feedback efficiency and its value is chosen to be 0.05 to match the normalisation of $\rm M_{BH}-\sigma$ relation with the current observation \citep{Dimatteoetal2005Natur}. For simplicity it is assumed that feedback energy is coupled to the surrounding medium isotropically. Mechanical energy coming out of the AGN in the form of a jet can be anisotropic which is not modeled in the simulation. However, mechanical feedback has been explored by various other groups in cosmological simulations \citep{Teyssier2002A&A, CattaneoTeyssier2007MNRAS, Teyssieretal2011MNRAS, Vazzaetal2013MNRAS, Baraietal2014MNRAS, Cieloetal2018MNRAS}.

The initial mass function of seed SMBH is unknown and hence to address the initial mass function the following prescription is adopted. The simulation runs an in-situ halo-finder algorithm and populates a halo with a seed SMBH when the mass of the halo exceeds a certain threshold. Halos are identified in this simulation with friends-of-friends (FOF) algorithm \citep{Davisetal1985ApJ}. In this algorithm particles are linked together if they are within a certain distance (linking length b) which is taken in K15 as $0.2\overline{l}$ ($\overline{l}$ being the mean particle separation). All the dark matter particles within this distance are taken altogether to form a halo. By this, all other particles, such as gas, stars and BHs also fall within the potential of their nearest DM halo. For identifying substructures within the halos the \begin{scriptsize}
SUBFIND\end{scriptsize} algorithm is used \citep{Springeletal2001MNRAS}. A seed black hole of mass $ \approx 10^{5} h^{-1}M_{\odot}$ is created within a halo of mass $ \approx 10^{10} h^{-1}M_{\odot}$ if the halo does not contain any black hole. The seed black hole is now allowed to grow through gas accretion and merger events.

We use the K15 data for two redshifts ($z=1.0$ and $z=0.1$) to study the redshift evolution of the central AGN and their host galaxies as a result of feedback on them. The K15 simulation was evolved upto $z=0$ but to compare our results with the D08 data we used the $z=1$ snapshot of K15. The larger boxsize of K15 comes with the following advantages that are crucial to our work: the number of higher mass halos and higher luminosity AGN increases with box size. These increase in numbers provide us the adequate statistics to probe the high end of the AGN luminosity as well as its host halo mass function. We further discuss this issue later. The simulation parameters for both the simulations are listed in Table \ref{table2}, where $N_{p}$ is the total number of gas + dark matter (DM) particles in the box, $m_{DM}$ and $m_{gas}$ are the mass resolutions for DM and gas particles respectively, $\varepsilon$ is the gravitational softening length and $z_{end}$ is the final redshift upto which the simulation evolves. It is to be noted that the without feedback cases are for the D4 simulation of \citet{Dimatteoetal2008ApJ}, which has a different resolution than the \citet{Khandaietal2015MNRAS} simulation.  We note that the D6 simulation of \citet{Dimatteoetal2008ApJ} includes feedback and has a similar resolution to that of \citet{Khandaietal2015MNRAS} although the boxsize of \citet{Khandaietal2015MNRAS} is bigger. \citet{Chatterjeeetal2008MNRAS} did a resolution study between the D4 and D6 simulations. It was deduced from that study that resolution effects should not affect the overall conclusion but the details of the temperature and density maps might be different for different resolutions (as has been seen with the case of Sunyaev-Zeldovich fluxes in \citealt{Chatterjeeetal2008MNRAS}). Unless specified our with feedback simulations are for K15 and the without feedback simulations are for D4-D08.
\begin{figure*}
\begin{center}
 \begin{tabular}{c}
     \includegraphics[width=8cm]{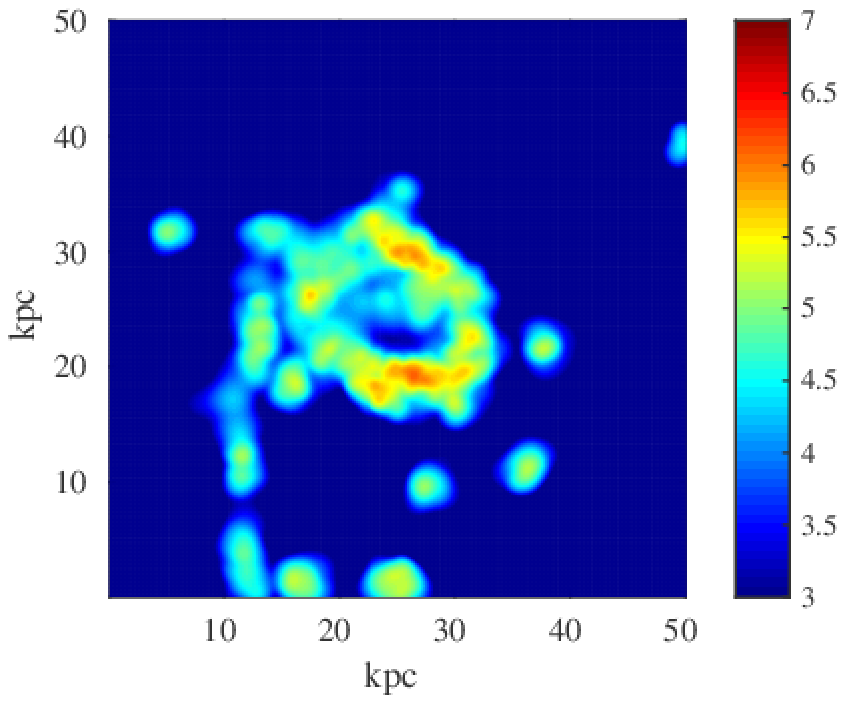}
      \includegraphics[width=8cm]{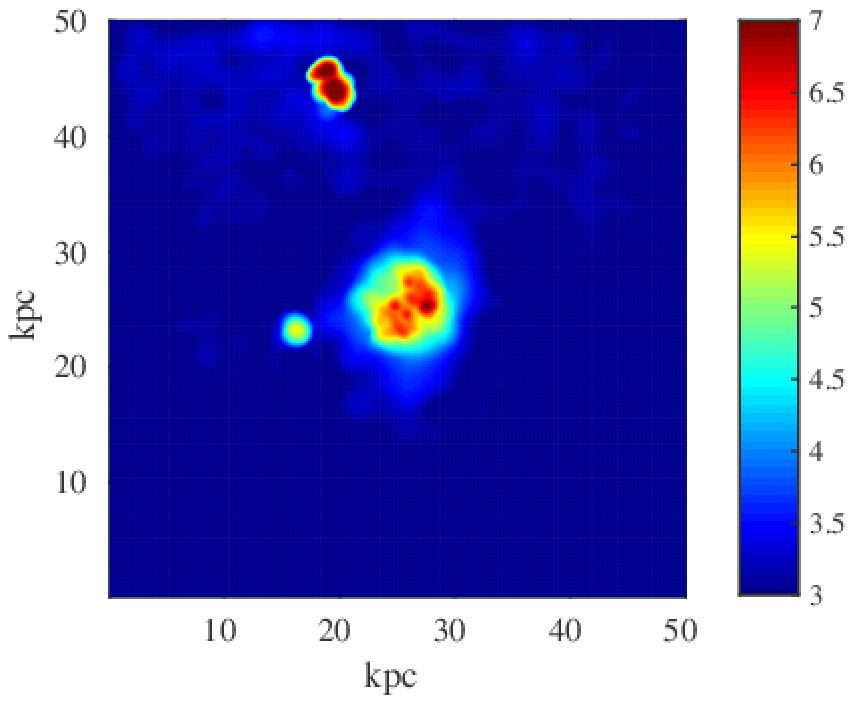}\\
      \includegraphics[width=8cm]{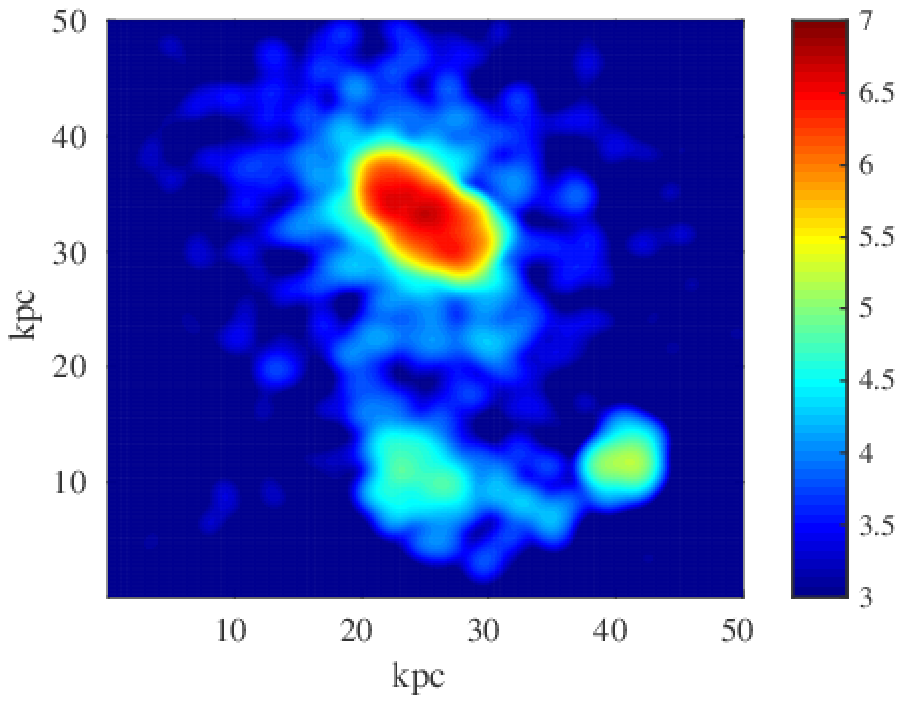} 
      \includegraphics[width=8cm]{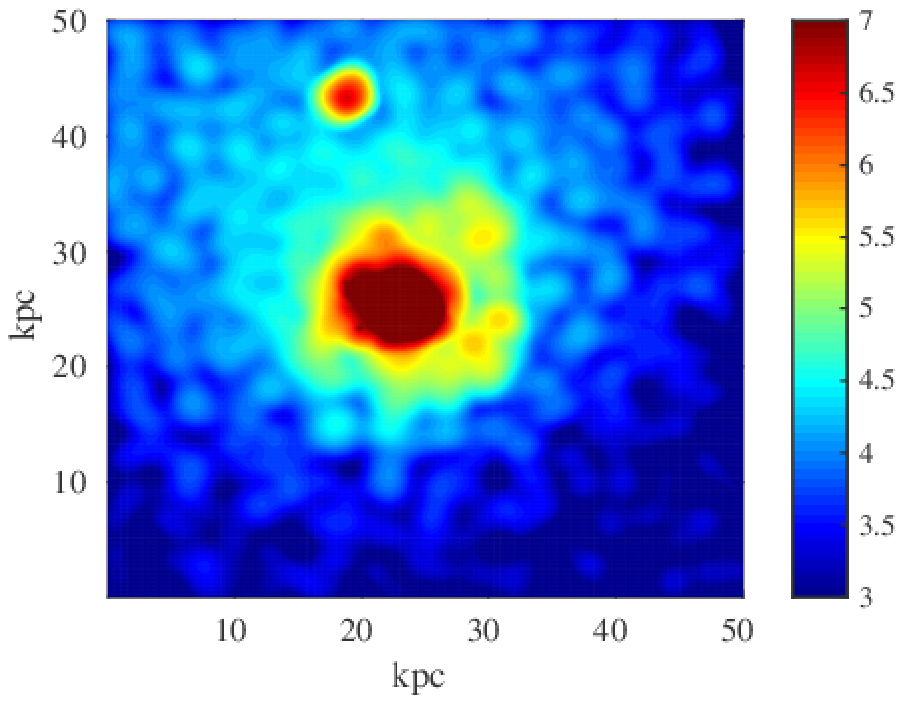}
\end{tabular}
 \caption{Density map in a region of radius 25 Kpc for different physical scenarios. Top panels show the density map for K15 simulation around two AGN having similar black hole mass $(\log_{10}(M_{BH})=7.7h^{-1}M_{\odot})$, residing in host halos of different masses. The density is shown in units of $M_{\odot}/kpc^{3}$. Bottom panels show the same maps, except for the case when AGN feedback is not present in D08 simulation. Top left panel : Density map for the BH hosted by a halo of mass $10^{12.2}h^{-1}M_{\odot}$. Top right panel : Density map for a BH residing in a host halo of mass $10^{13.9}h^{-1}M_{\odot}$. It is clearly seen that at higher mass halo density of gas is high at the centre. Bottom left panel : Density map inside a halo of mass $10^{12.1}h^{-1}M_{\odot}$. Bottom right panel : Density map inside a halo of mass $10^{13.4}h^{-1}M_{\odot}$. In the lower panels too we see higher density of the diffuse gas for higher mass halos. However we can see a clear enhancement of the density when feedback is not present at both halo mass scales. This we argue is due to the effect of AGN feedback as discussed in \S 3 and \S 4. }
  \label{fig:halo_comparison}
\end{center}
 \end{figure*}

For the current study we select black holes which have mass $\geq 10^{7}h^{-1}M_{\odot}$. As already discussed, the simulation inserts seed BHs of mass $10^{5}h^{-1}M_{\odot}$ when halos have a mass greater than $10^{10} h^{-1}M_{\odot}$. These black holes then grow by accretion or merger. We thus prefer to limit our analysis of black holes with mass well above the seed black hole mass to minimize the numerical noise. As a result of this, host halos of the lower mass black holes are automatically being discarded. Thus we are left with 2730 and 3702 black holes at $z=1$ and $z=0.1$ respectively. One of the goals of this study is to understand the role of AGN on structure formation and hence a BH mass cut-off comes as a better choice since to first order, luminosity also scales as black hole mass.  Also it is seen that $\rm M_{BH}$ and $\rm M_{halo}$ are correlated but the onset of physical processes like AGN feedback alter these correlations and they can be studied in an unbiased way if we go to a mass limit away from the seed black hole mass. We checked that the effect of AGN feedback is evident at the higher end of both the halo mass and the black hole mass functions and hence the lower cut-off on black hole mass or the halo mass will not alter our results.

In our work we study the scaling relations between the properties of the SMBH and their host halos. We also investigate the redshift evolution of those scaling relations. We propose the effect of AGN feedback on its surrounding medium as to explain these scaling relations and a probe to study the interaction of the SMBH with its environment. To do this we focus on the diffuse gas as well as total stellar mass inside the halos. Density and temperature distributions of gas surrounding the AGN clearly reveal the effect of AGN feedback on its surrounding environment. We present these results in the following section.  
 
\begin{figure*}
\begin{center}
 \begin{tabular}{c}
     \includegraphics[width=8cm]{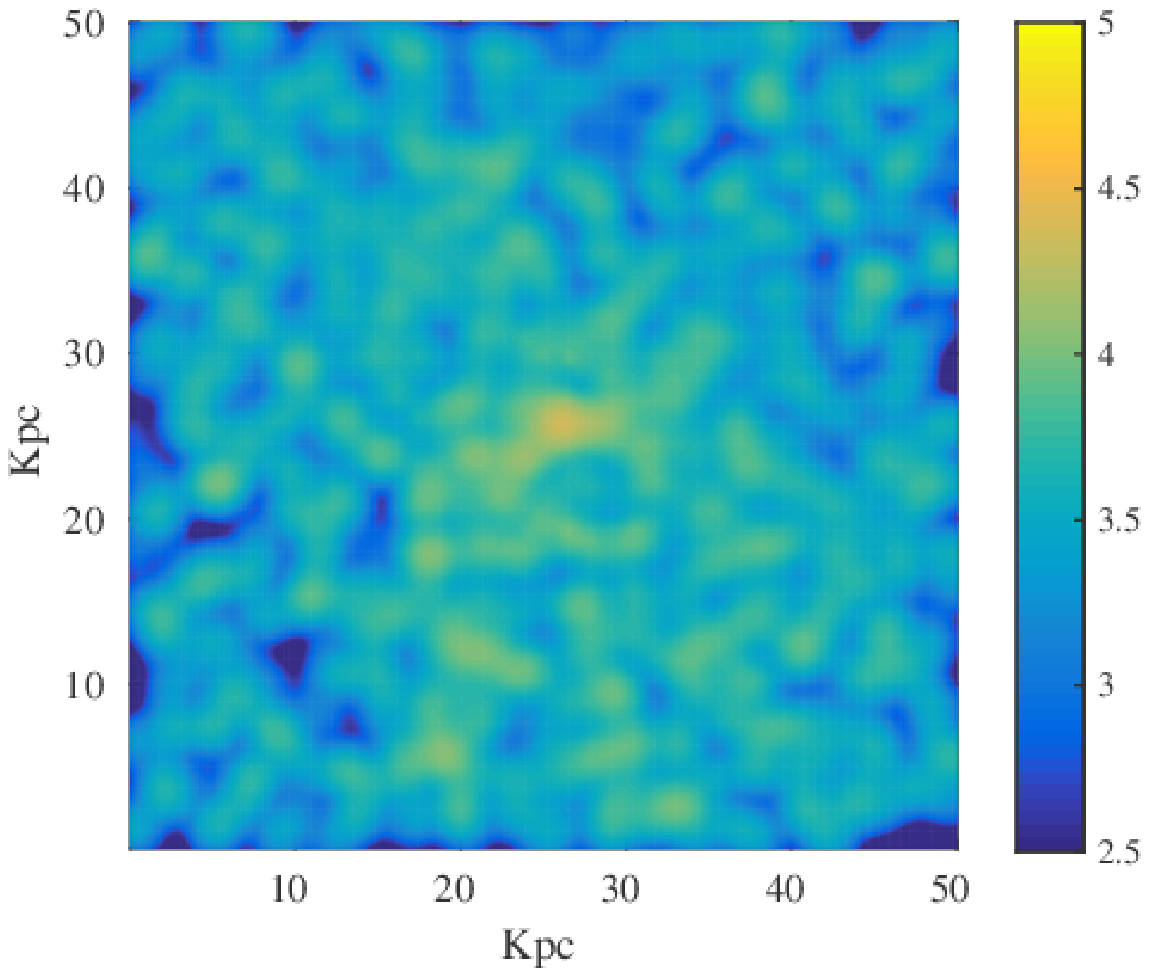}
      \includegraphics[width=8cm]{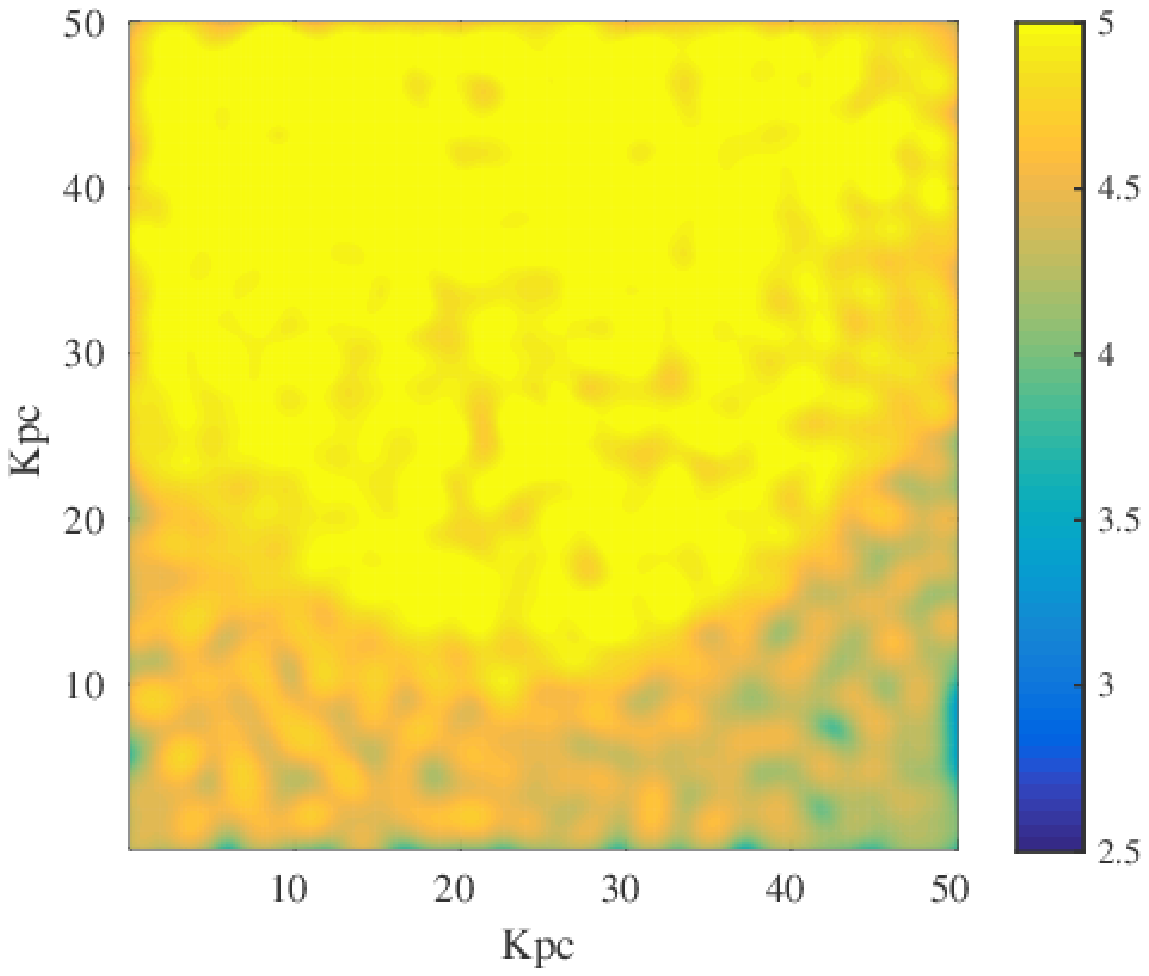}\\
      \includegraphics[width=7.5cm]{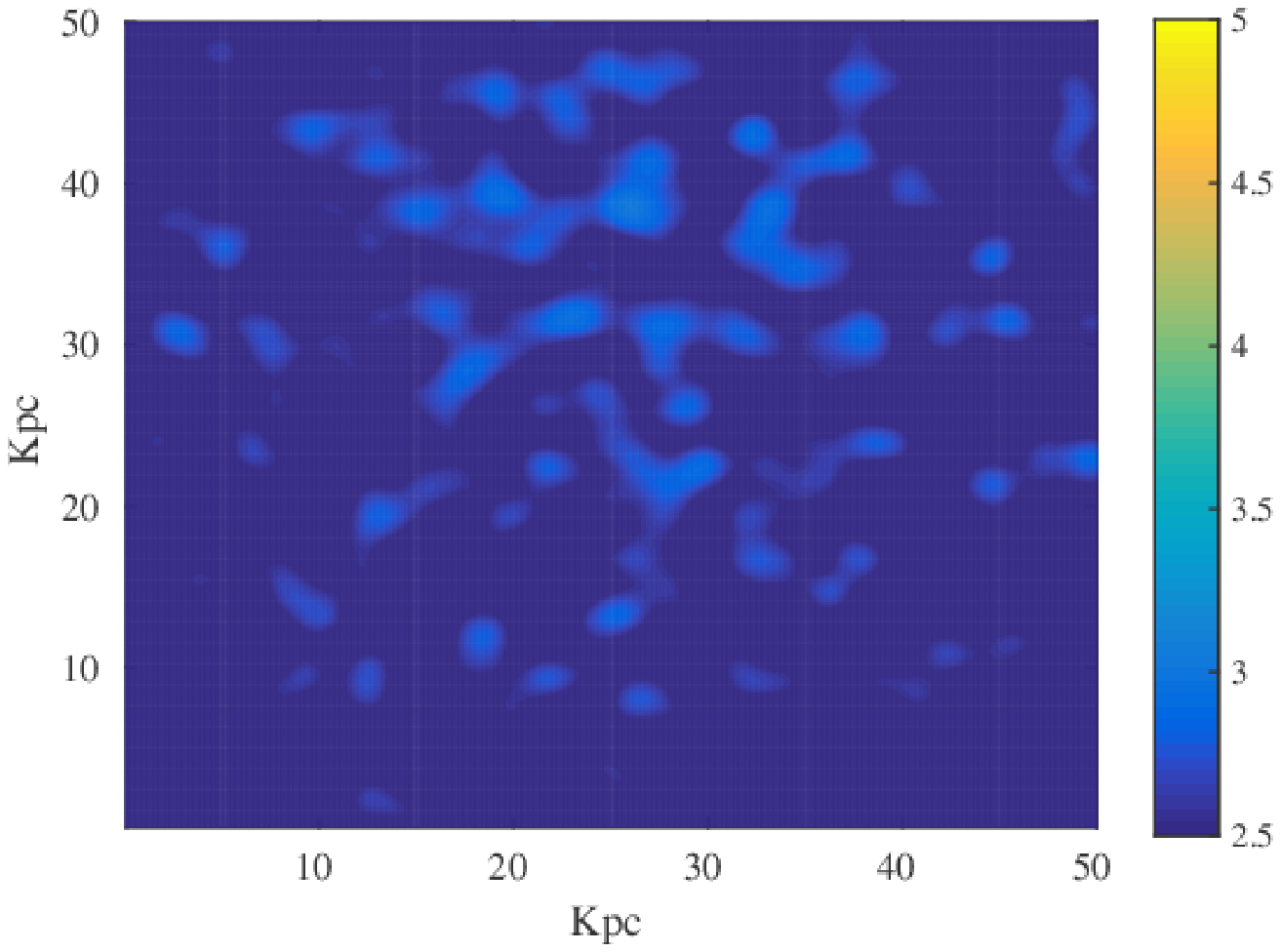} 
      \includegraphics[width=7.5cm]{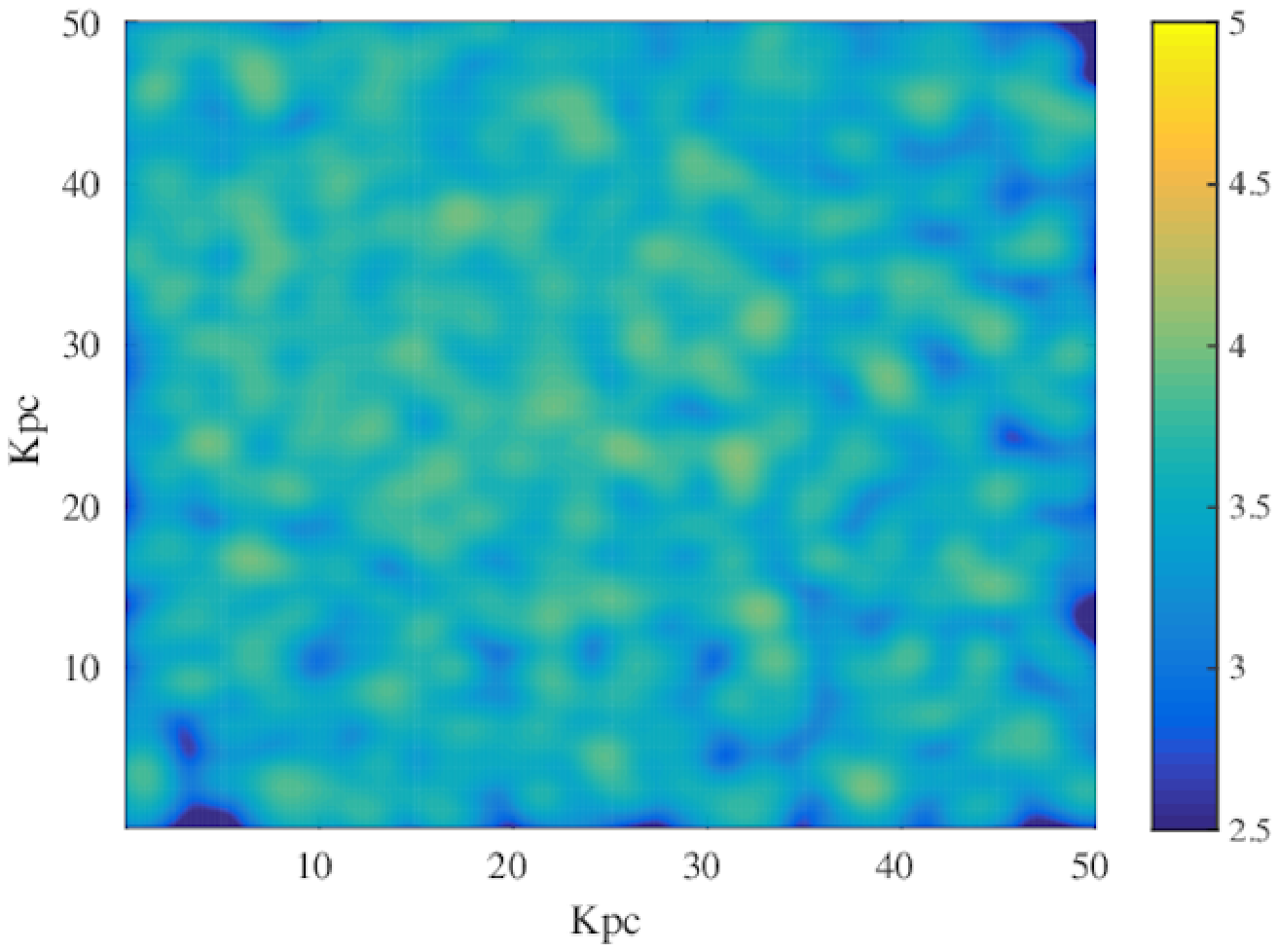}
\end{tabular}
 \caption{Temperature map in a region of radius 25 Kpc for different physical scenarios. Top panels show the temperature maps for K15 simulation around two AGN having similar black hole mass $(\log_{10}(M_{BH})=7.7h^{-1}M_{\odot})$, residing in host halos of different masses. Unit of temperature is Kelvin. Bottom panels show the same maps, except for the case when AGN feedback is not present in case of D08 simulation. Top left panel : Temperature map of the diffuse gas around the BH hosted by a halo of mass $10^{12.2}h^{-1}M_{\odot}$. Top right panel : Temperature map of surrounding gas of BH residing in a host halo of mass $10^{13.9}h^{-1}M_{\odot}$. It is clearly seen that at higher mass halo temperature is high at the centre. Bottom left panel : Temperature map inside a halo of mass $10^{12.1}h^{-1}M_{\odot}$. Bottom right panel : Temperature map inside a halo of mass $10^{13.4}h^{-1}M_{\odot}$. In the lower panels too we see higher temperature for higher mass halos. However we can see a clear enhancement of temperature when feedback is present at both halo mass scales. This we argue is due to the effect of AGN feedback as discussed in \S 3 and \S 4. }
  \label{fig:halo_comparison_temp}
\end{center}
 \end{figure*}

\section{Results}
Fig. \ref{fig:scaling2} shows the correlation between the properties of the central BH with their host halos at $z=1$ and $z=0.1$ for K15. The BH population has been categorized into central and satellite BH at both the redshifts and the correlation of their masses with their host halo masses are shown in the left panel. The number of central BH at $z=1$ and $z=0.1$ are 1956 and 2514 respectively while the number of satellite BH are 774 and 1188 respectively. It is observed that the central BH mass is increasing with host halo mass at both $z=1$ and $z=0.1$ and we do not see any significant redshift evolution of this correlation. The satellite BH mass slightly increases and then gets saturated at the higher mass end of the host halos. Individual history of the most massive SMBH has been discussed in K15 in $\S 8$. They have found that gas accretion dominates the growth of the most massive BH with the final mass being $10^{9} - 10^{10}M_{\odot}$ at $z=0$. It is reported in K15 that only 20\% of the total mass of the BH is gained by the merger of which 75\% of the mass is gained below $z=1$.

The error on the data points are standard error, calculated as $\sigma / \sqrt{N}$ , where $\sigma$ is the standard deviation and $N$ is the total number of data points in a particular mass bin. We emphasize that the error on the `mean' represents the reliability of the `mean' as a representative number of the binned data.  Hence standard error on the mean is used to signify the statistical power of the averaging.

The right panel depicts the variation of bolometric luminosity with the host halo mass of BH. Average bolometric luminosity for each halo mass bin is shown in the figure. We note that the luminosities used in this study refer to an instantaneous value. The time-averaged luminosities can also be considered. However when we consider a statistical sample, we emphasize that the time-averaged luminosity will be equivalent to the average instantaneous luminosity of many black holes (see discussions in \citealt{Chatterjeeetal2012MNRAS}). We thus feel confident that the characterization of instantaneous luminosities will be adequate in comparing our theoretical work with observations. We observe that with the increase of the halo mass, bolometric luminosty of the central AGN increases at both redshifts. But the satellite black hole luminosity is decreasing with the host halo mass. The downsizing observed in the $\rm L_{Bol}-M_{halo}$ relation might be a hallmark of the effect of AGN feedback in the simulation. It is likely that the feedback from the central BH is regulating the growth of the satellite BH (see \citealt{Chatterjeeetal2012MNRAS} for discussions). 

\begin{figure*}
\begin{center}
 \begin{tabular}{c}
     \includegraphics[width=8cm]{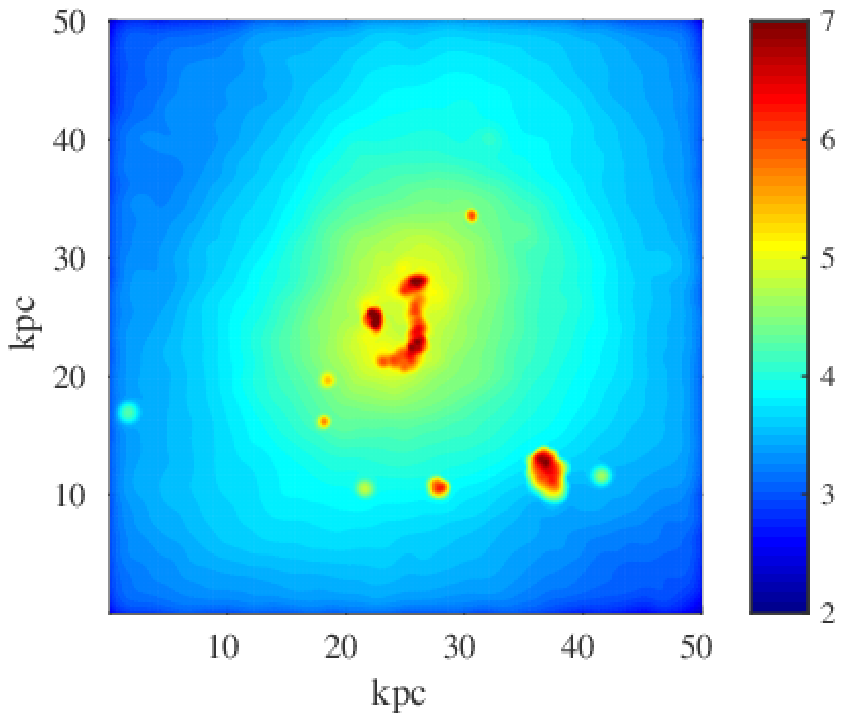}
      \includegraphics[width=8cm]{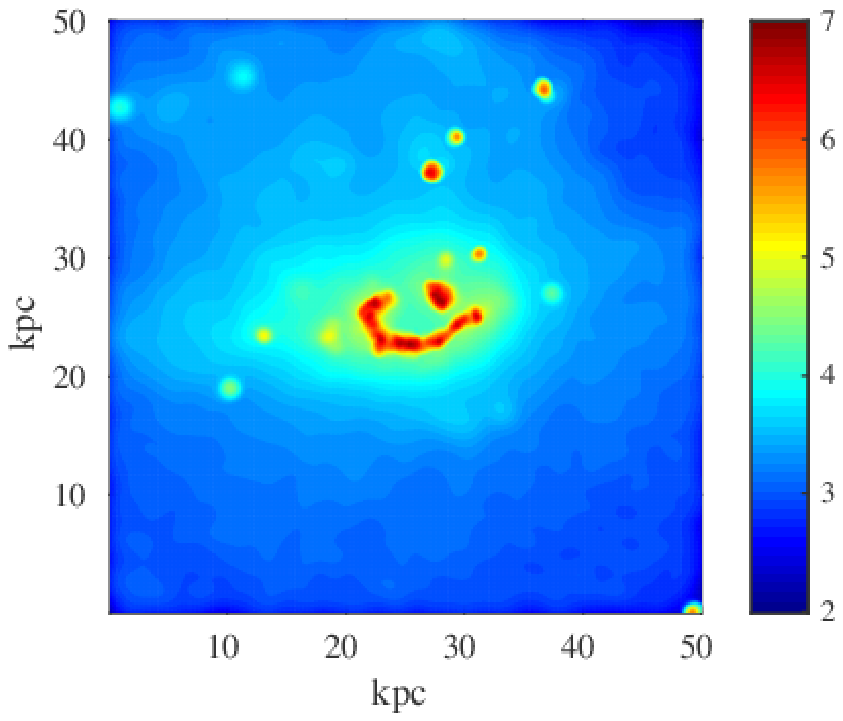}\\
      \includegraphics[width=8cm]{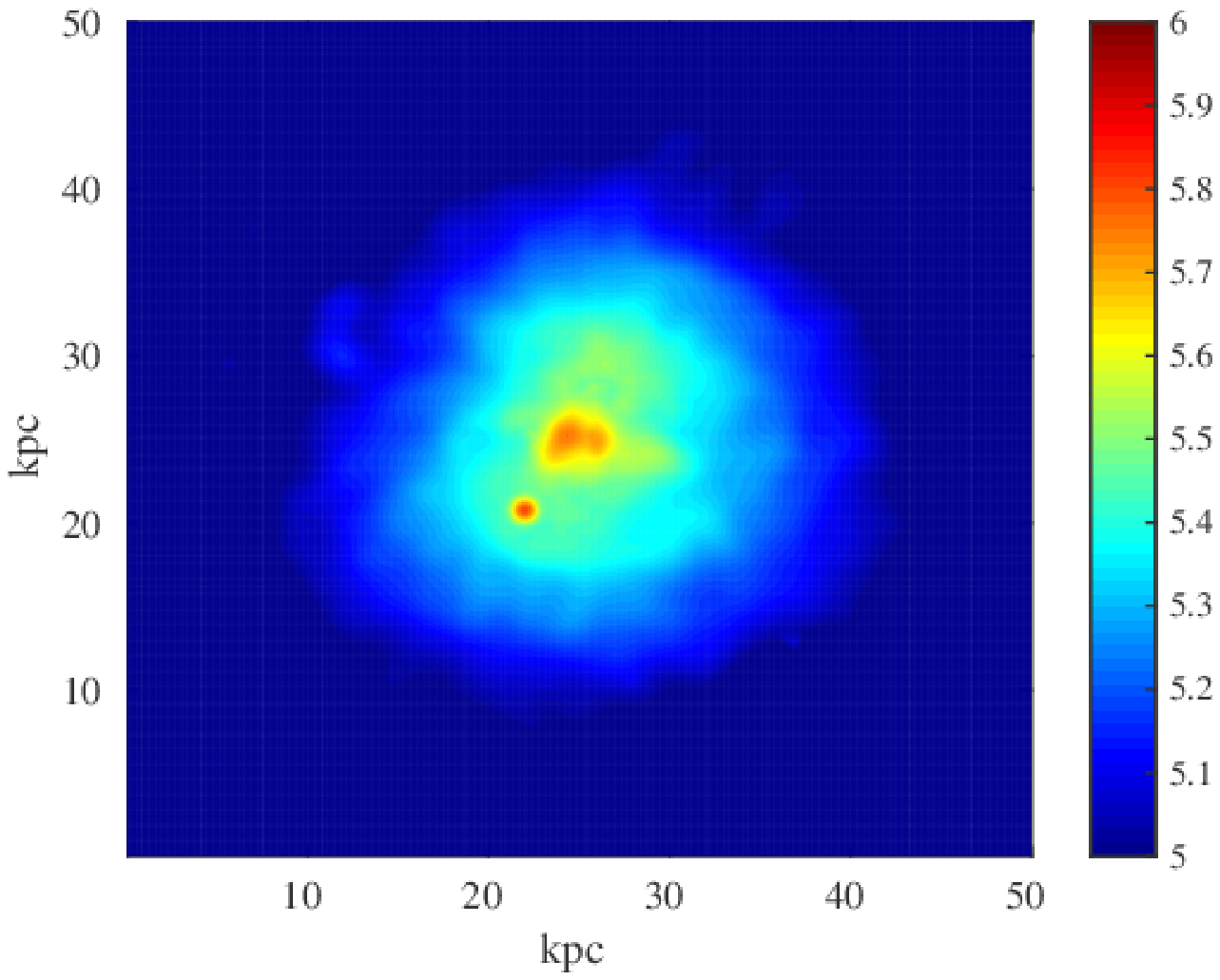}
      \includegraphics[width=8cm]{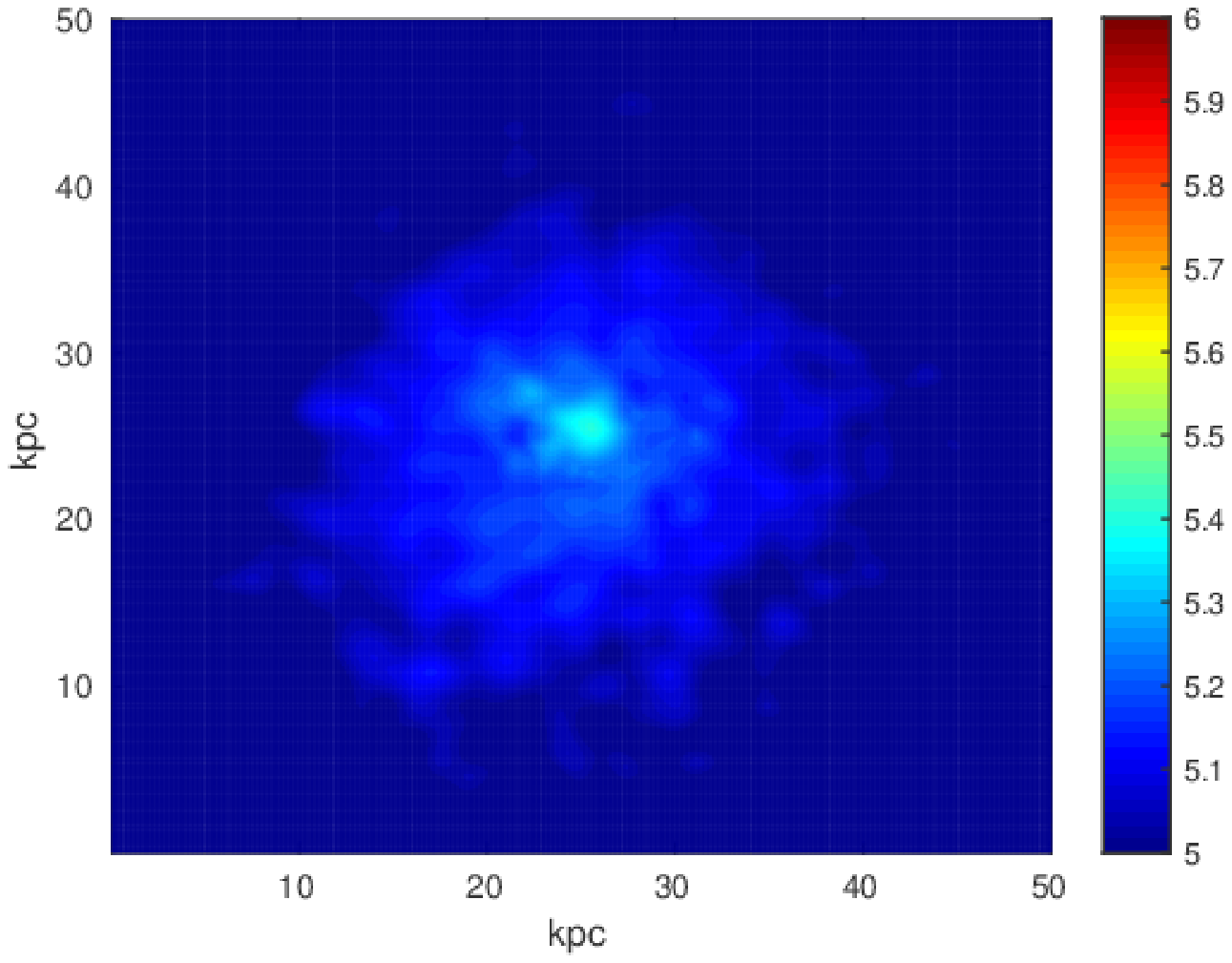} 
\end{tabular}
 \caption{Density and temperature maps in a region of radius 25 Kpc around two AGN of similar mass in different halo environment at $z=0.1$. Top panels show the density maps for K15 simulation around two AGN having similar black hole mass $(\log_{10}(M_{BH})=8.2h^{-1}M_{\odot})$, residing in host halos of different masses. The density is shown in units of $M_{\odot}/kpc^{3}$. Bottom panels show the temperature maps for the similar situation. Colorbar shows the temperature in Kelvin. Top left panel : Density map of the diffuse gas around the BH hosted by a halo of mass $10^{12.8}h^{-1}M_{\odot}$. Top right panel : Density map of surrounding gas of BH residing in a host halo of mass $10^{13.8}h^{-1}M_{\odot}$. Bottom left panel : Temperature map inside a halo of mass $10^{12.8}h^{-1}M_{\odot}$. Bottom right panel : Temperature map inside a halo of mass $10^{13.8}h^{-1}M_{\odot}$. Effect of AGN feedback can be clearly seen from the density maps. }
  \label{fig:halo_comparison_0.1}
\end{center}
 \end{figure*}

To further examine Fig. \ref{fig:scaling2} we plot the conditional $\rm M_{BH}-L_{Bol}$ distributions of central and satellite black holes for two redshifts in Fig. \ref{fig:combine} and Fig. \ref{fig:combine0.1}. The luminosities represent the average in each black hole mass bin. We note that the luminosity (accretion rate) is almost constant with black hole mass in the lower halo mass bins. Central black hole luminosity is almost constant or increase very slightly with black hole mass in all the halo mass bins at redshift 1.0 and 0.1. At $z=1.0$ satellite BH luminosity remains unchanged with the BH mass upto a halo mass scale $\approx 10^{13.2}h^{-1}M_{\odot}$. However, above this halo mass range the distribution of luminosities gets wider while at the highest mass bin we do see the satellite AGN with low luminosities. This feature can be also seen at $z=0.1$. At this redshift the luminosity of the satellite AGN spreads above a halo mass bin of $\approx 10^{12.9}h^{-1}M_{\odot}$. The emergence of lower luminosity satellite AGN in higher mass halos might be a signature of AGN feedback where the supply of cold gas gets cut-off due to the additional heating from the central engine.

To explain the result, we show the distribution of the density and temperature of the gas adjacent to the AGN for cases with and without feedback and for two different host halo masses at $z=1.0$. To differentiate the effect of the halo (gravitational effect) with that of the AGN feedback (non-gravitational effect) on the surrounding diffused gas we take two black holes at $z=1$ with identical black hole mass but different host halo masses, and study the density and temperature distribution of their surrounding hot gas. We compute the smoothed temperature and density using the cubic spline kernel.

\begin{figure*}
 \begin{tabular}{c}
     \includegraphics[width=9cm]{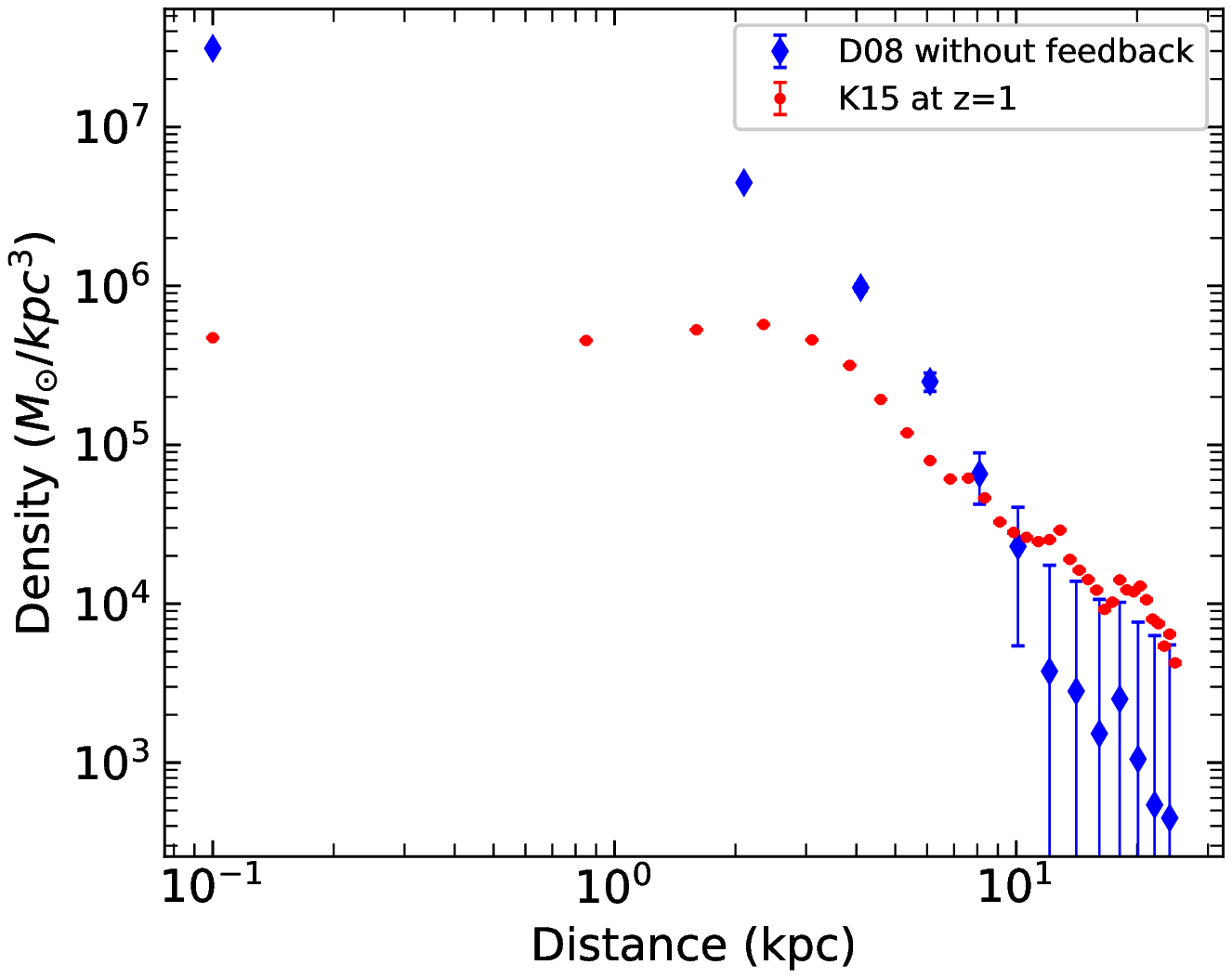}
      \includegraphics[width=8.2cm]{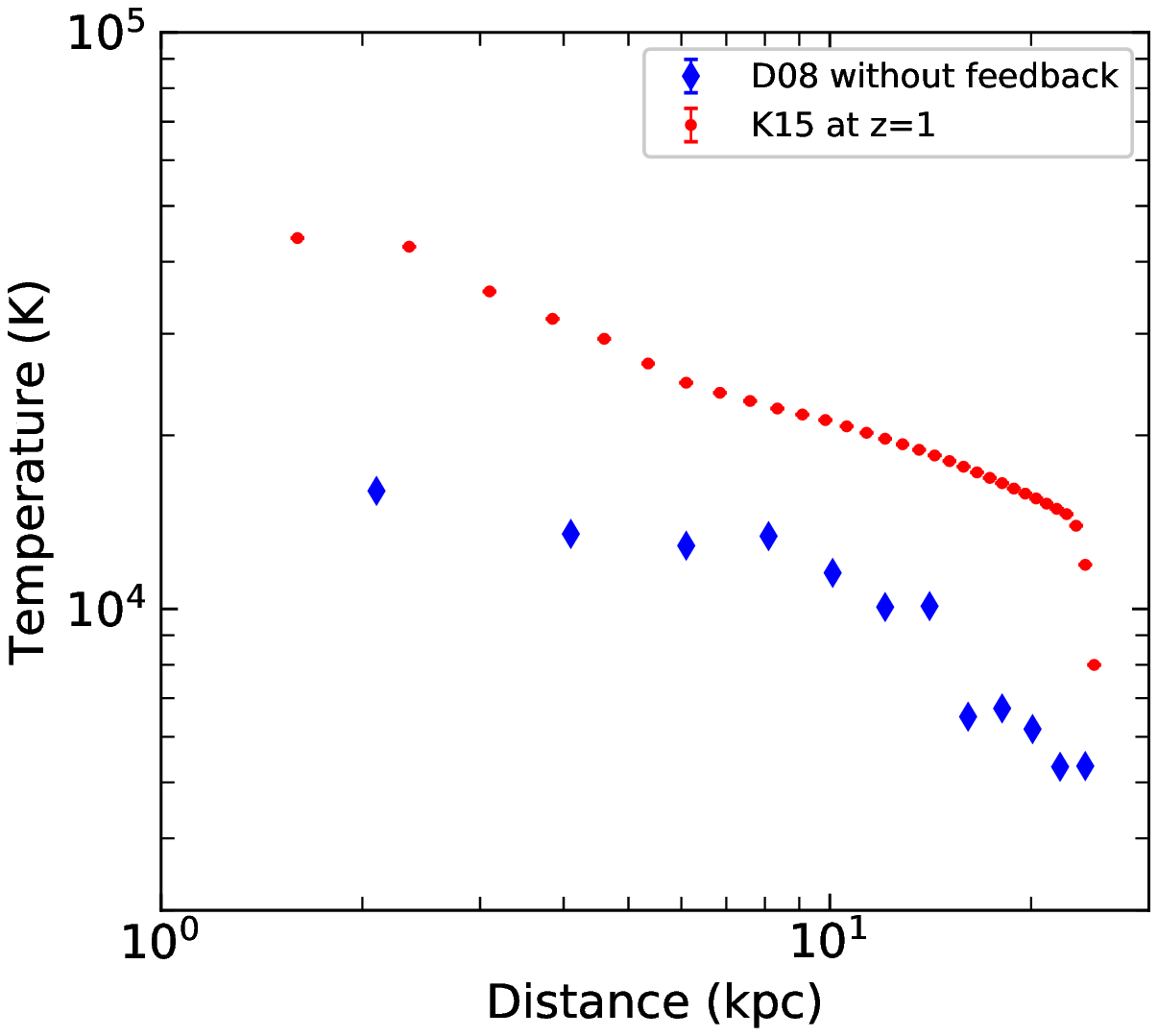}
     \end{tabular}
 \caption{Left panel : Radial profile of the stacked density in presence and absence of AGN feedback inside a region of radius 25 kpc at $z=1$. Blue diamond represents the density profile when AGN feedback is absent in D08. Red dot shows the scenario when AGN feedback is present in K15. A significant reduction of density at the central region in the presence of AGN feedback supports our claim of displacement of hot gas as a result of AGN feedback.  Right panel : Stacked radial profile of the temperature in presence and absence of AGN feedback inside a region of radius 25 kpc at $z=1$. Blue diamond signs are representative of the temperature profile when AGN feedback is absent in D08. Red dots exhibit the condition when AGN feedback is present in K15. Error bars on the data points in both the panels are the standard errors as discussed in \S 3 divided by the square root of the number of stacked maps to include the stacking error. A clear enhancement of temperature at the central region in the presence of AGN feedback supports our claim of heating the surrounding gas as a result of AGN feedback. See \S 3 and \S 4 for more discussions.}
  \label{fig:profile}
 \end{figure*}

The results are shown in Fig. \ref{fig:halo_comparison} and Fig. \ref{fig:halo_comparison_temp}. Both black holes have mass $7.7h^{-1}M_{\odot}$ in logarithmic scale. Both the density and temperature maps in the top right panel is shown for a BH of host halo mass $7.9 \times 10^{13}h^{-1}M_{\odot}$ and the top left panel shows the map around a BH residing in a halo of mass $1.6 \times 10^{12}h^{-1}M_{\odot}$. To understand the effect of feedback, we constructed two more density and temperature maps using no feedback data (bottom panel). Bottom left panel of Fig. \ref{fig:halo_comparison} and Fig. \ref{fig:halo_comparison_temp} shows density and temperature map respectively inside a lower mass halo $(\rm M_{halo} = 1.3 \times 10^{12} h^{-1} M_{\odot})$ and bottom right panel is representing the map inside a higher mass halo $(\rm M_{halo} = 2.5 \times 10^{13} h^{-1}M_{\odot})$. It is clearly seen from Fig. \ref{fig:halo_comparison} and Fig. \ref{fig:halo_comparison_temp} that average density and temperature is higher in the higher halo mass in both the cases when AGN feedback is present and absent in the simulation. But the presence of feedback decreases density at the centre significantly at both halo mass scales whereas temperature increases with the feedback. If it were alone the effect of gravitational potential, then we should get the same results for the presence and absence of AGN feedback. It is to be noted that in Fig. \ref{fig:halo_comparison_temp}, temperature maps look somewhat grainy. We understand that this is related to the spatial scale of the maps. To investigate this issue further, we have checked the distribution at a larger scale (200kpc) and found that the maps are smooth, representing the SPH particle field. We also checked that the scaling relations at this larger spatial scale, and confirmed that statistically the relations stay the same without altering any of the conclusions.

To study the effect of AGN feedback at $z=0.1$, we made similar temperature and density maps around two black holes sitting in two different environment - one inside a low mass halo and another inside a host halo having higher mass. This is shown in Fig. \ref{fig:halo_comparison_0.1}. Both the black holes have very similar masses of $10^{8.2}h^{-1}M_{\odot}$. Left panels show the result for the halo of mass $10^{12.8}h^{-1}M_{\odot}$. Right panels show the results of halo having higher mass $10^{13.8}h^{-1}M_{\odot}$. Top panels show the density distribution inside a region of radius 25 kpc for two AGN from K15 simulation in two different halo mass. Temperature distribution of those two AGN is shown in the bottom panels. We also note that the bolometric luminosities of two AGN residing at lower and higher mass halo are $10^{43}$ and $10^{44}$ erg/sec respectively. So, the AGN in the lower mass host halo is more active and this is reflected in the temperature map where we can see an enhancement of temperature at the centre of the bottom left map when compared with the bottom right one. Effect of feedback is also evident from both the density maps where it is seen that the gas density is lower at the central region of the maps. However, we note that due to the limitations of both D08 and K15 simulation it is not possible to compare the results of $z=0.1$ with the scenario when feedback is absent. However we argue that the effect of enhancement of temperature and suppression of density due to feedback effects are ubiquitous at all redshifts. 

Now, to check the effect of feedback statistically on the density and temperature of the gas inside the halo, we have obtained the stacked radial profiles of density and temperature in presence and absence of AGN feedback at $z=1$. The results are shown in Fig. \ref{fig:profile}. Here the error bars on the data points are the standard errors as discussed in \S 3 divided by the square root of the total number of stacked maps in order to take into account the stacking error. Left panel of this figure shows the radial profile of the stacked density inside 25kpc radius region. Blue diamonds represent D08 without feedback density profile while red dot shows the density profile for K15 simulation. We can see a decrement of gas density at the central region in the presence of AGN feedback supporting our claim that feedback energy drives the hot gas outward, creating the under dense region surrounding the AGN. Slight enhancement of density in presence of feedback is seen over the without feedback simulation at a radii $>10$ kpc although this is within the error bar. However, we explain that this overdensity is probably coming from the accumulation of gas that has being pushed out of the central region due to feedback. Similar feature is also found in the radial profile of the X-ray flux \citep{Mukherjeeetal2019ApJ} as the main X-ray  emitting mechanism from clusters is thermal bremsstrahlung. So, it is expected that the density profile will follow the X-ray profile. Right panel of Fig. \ref{fig:profile} represents the stacked temperature profile. Blue diamonds and red dots show the stacked temperature for D08 without feedback and K15 simulation respectively. The clear enhancement of temperature in the presence of AGN feedback is in accordance with our claim that the gas surrounding AGN is heated by the feedback energy.

\begin{figure}
\includegraphics[width=8.5cm]{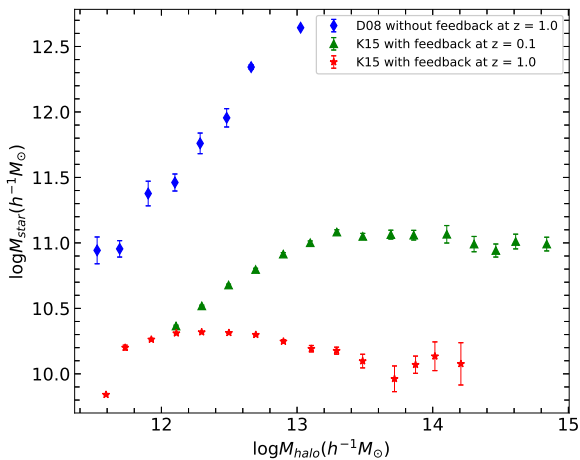}
\caption{Variation of stellar mass (see definition of stellar mass in \S 2) with the corresponding halo mass bin. Blue diamonds represent the case when AGN feedback is absent in D08 simulation at $z=1$. Red stars and green triangles are representative of K15 simulation at $z=1$ and $z=0.1$ respectively. A clear deficit of stellar mass is seen in the presence of AGN feedback in the same halo mass bin when compared with the simulation where AGN feedback is absent. The results are discussed in \S 3 and \S 4}
\label{fig:star}
 \end{figure}

Finally, to check the effect of AGN feedback from a different perspective, we plot the stellar mass with the corresponding halo mass. We define stellar mass as the total mass of the stars inside a radius which is equal to $0.2R_{200}$. $R_{200}$ is the usual radius where the density of the halo becomes 200 times the critical density. This is shown in Fig. \ref{fig:star}. Here blue diamonds represent the case when AGN feedback is absent in D08 simulation at $z=1.0$. Red stars and green triangles are for K15 simulation at $z=1.0$ and $z=0.1$ respectively. It is clearly seen from the figure that the presence of feedback reduces the stellar mass when compared with the case of `without feedback'. Also, it is seen that the stellar mass is higher at $z=0.1$ compared to $z=1.0$. Here we note that the definition of stellar mass used in this paper may be different from the other groups. For example, \citet{Khandaietal2015MNRAS} define galaxies as the subhaloes consisting greater than 100 dark matter particles and their stellar mass is represented by the mass of the subhalos. In \citet{Torreyetal2014MNRAS} stellar mass is defined as the total mass of the stars within twice the half-mass radius. We wish to carefully note this difference and emphasize the fact that feedback from AGN shuts down star-formation in its vicinity by limiting the availability of cold gas. 

\section{Discussion}

In this work we study the global role of AGN feedback on the accretion of the AGN itself as well as its impact on large scale environments of AGN, using a cosmological volume simulation. In the left panel of Fig. \ref{fig:scaling2} we observe a strong correlation between the mass of the central SMBH with their host dark matter halos, consistent with other observation and simulation results \citep[e.g.,][]{Ferrarese2002ApJ, Baesetal2003MNRAS, Colbergetal2008MNRAS, Fillouxetal2010IJMPD, BoothSchaye2010MNRAS, Sabraetal2015ApJ, Khandaietal2015MNRAS}. We see that at both redshifts 1.0 and 0.1, central BH mass is increasing with its host halo mass while the satellite BH mass is increasing weakly while saturating at the higher mass end. However, no redshift evolution of this correlation is observed.

The right panel of Fig.\ 1 shows a downsizing of satellite AGN luminosity with halo mass at $z=1.0$ and $z=0.1$, a result that has been observed in clustering studies of AGN \citep[e.g.,][]{Richardsonetal2012ApJ, Richardsonetal2013ApJ}. However, the bolometric luminosities of the central AGN increases with the host halo mass. It is possible that the accretion and feedback from the central AGN affects the evolution of the satellites. \citet{Chatterjeeetal2012MNRAS} tried to investigate the effect of feedback from the central AGN on the satellite AGN but they did not find any significant effect. But we note that the studies of \citet{Chatterjeeetal2012MNRAS} were confined to halo mass scales of $10^{13}M_{\odot}$. Our results too do not exhibit a substantial difference till a mass scale of $10^{13}M_{\odot}$ but we find that the effect becomes significant above this halo mass scale. \citet{Richardsonetal2013ApJ} showed that bright optically selected quasars with higher bolometric luminosity reside in lower mass halos compared to X-ray AGN with less luminosity. Some authors claim that the depletion of cold gas inside higher mass halos due to shock heating is responsible for shutting down black hole accretion \citep[e.g.][]{Sijackietal2007MNRAS, Dimatteoetal2012ApJ, Gasparietal2013AN}. Another possibility that exists in the literature is `AGN feedback.' In this study we addressed this question by looking at the density and temperature of the gas in the vicinity of the black hole.

In a given cluster/galaxy we can consider two effects: gas falling into the gravitational potential well of the cluster gets heated, and the central SMBH displacing the gas via mechanical or radiative feedback. If the effect of halo dominates over feedback then an increase in density of the diffuse gas is expected in the central region of the halo and BH mass will be correlated with halo mass. But these correlations can get altered if additional heating or cooling mechanisms are present. In Fig. \ref{fig:scaling2} we observe a downsizing of the BH luminosity. We argue that the feedback effect from AGN is at play here. Outflow from the AGN will displace the hot gas outward, thereby reducing the average density. So, the supply of cold gas gets cut inside the higher mass halo which results in the suppression of the BH mass. Another possibility for this effect would be the shock heating inside the halo itself.

To disentangle the effect of feedback and halo potential on the diffuse gas, we compute the density and temperature maps around two equally massive BHs (Fig.\ref{fig:halo_comparison} and Fig. \ref{fig:halo_comparison_temp} respectively) and compared them with the scenario when feedback is absent (respectively Fig. \ref{fig:halo_comparison} and Fig. \ref{fig:halo_comparison_temp} bottom panels). In Fig. \ref{fig:halo_comparison}, left panels are density maps around AGN residing at higher mass halos and right panels are density maps of diffused gas around AGN at lower mass halos. We clearly see an excess of the density in the left panels which represent higher mass halos in both cases. However, when we compare the `with feedback' case with the `no feedback' case, we observe a deficit in density at both halo masses.

In Fig. \ref{fig:halo_comparison_temp} we compute the average smoothed temperature of the gas in the central region of the halo for all four cases and we find that with the presence of AGN feedback the temperature of the gas is higher at both halo mass scales as has been reported previously by \citet{Chatterjeeetal2008MNRAS}. We thus propose that AGN feedback is a significant mechanism of heating and depletion of the cold gas inside the higher mass halos, thus resulting in the observed downsizing in Fig. \ref{fig:scaling2} (AGN luminosity with halo mass). 
This result is also supported by Fig. \ref{fig:combine} \& \ref{fig:combine0.1}, where we have plotted the variation of bolometric luminosity of central and satellite AGN with their mass at different host halo mass bins at $z=1$ and $z=0.1$ respectively. We see a spread in the bolometric luminosity of satellite AGN in the higher mass host halo bins with very low luminosity AGN.

In Fig \ref{fig:halo_comparison_0.1}., we examine the effect of AGN and the host halo potential on the hot gas at $z=0.1$. From the density map the signature of AGN feedback is evident at the central region where the gas being pushed outward, resulting a deficit in the density map. However, from the temperature map, we can see that the temperature of the gas is higher at the low halo mass. Here we note that the bolometric luminosity of the BH inside the low mass halo is higher. Hence we can conclude that the more active AGN energizing its surrounding medium more, causing an enhacement of temperature.

Finally in Fig \ref{fig:profile}., we investigate the effect of AGN feedback on the density and temperature of the gas statistically by computing their stacked profile. It is evident from this figure that density of the gas decreases statistically in the presence of AGN feedback. Moreover, right panel of the figure shows that presence of AGN feedback is enhancing the temperature of central region of the halo statistically. The overall enhancement of temperature and decrement of density of the gas inside all the halo supports our claim of feedback activity. In addition to temperature and density we study the correlation between their stellar mass with the corresponding halo mass in the presence and absence of AGN feedback (Fig. \ref{fig:star}). In this figure, we see that the stellar mass is significantly low in the presence of AGN feedback when compared with the case of absence of feedback. We can explain that AGN feedback heats the surrounding gas significantly that the availability of cold gas gets depleted thereby reducing the stellar mass.

In recent studies involving X-ray stacking analysis of AGN at moderately high redshift (z $\approx 0.6$) \citet{Chatterjeeetal2015PASP} and \citet{Mukherjeeetal2019ApJ} showed that AGN at large have lower X-ray fluxes in the central region compared to the X-ray flux of galaxies without AGN. They propose a promising technique in studying the X-ray gas in higher redshift galaxies for quantifying AGN feedback. In this work we study the effect of AGN feedback on temperature, density and stellar mass and examine their global effects. In a follow up paper, using K15 and D08 data (Kar Chowdhury et al. 2019B, in preparation) we perform a detailed X-ray analysis of the intra-cluster medium and compare our results with the studies of \citet{Chatterjeeetal2015PASP} and \citet{Mukherjeeetal2019ApJ}. 

Finally, we note that the simulation used for this work has several limitations. The accretion onto the BH is based on a simplistic spherical accretion model. Spatial resolution of the simulation limits to include all the precise relativistic accretion flow in the simulation. Further, the energy due to feedback from the black hole is assumed to be distributed isotropically among the gas particles surrounding them in a region within the smoothing length. Momentum driven outflow from the AGN has not been taken into account in the simulation. But we note that as the effects of BH accretion on the resolved large scale environment match with the observations, Bondi model can be used to study the effect of AGN on the cosmological scale \citep{Dimatteoetal2008ApJ}. Also it has been carefully noted that the effect of linking the black hole with its surrounding medium is independent of the coupling model as long as the coupling scale and time is small compared to the scale and dynamical time of the host galaxy \citep{Dimatteoetal2008ApJ}.

A plethora of work has been done by different groups to understand the growth of AGN and its effect on the surrounding medium using cosmological simulation and correlate the results with different observables. \citet{LeBrunetal2014MNRAS} ran cosmological hydrodynamic simulation cosmo-OWLS to examine different scaling relations between various properties of galaxy groups and clusters and compared it to observations. They found that the density profile is lower at the central region in presence of strong AGN feedback activity which agrees well with our result (left panel of Fig. \ref{fig:profile}). Also, the $\rm M_{BH}-M_{halo}$ relation obtained by them qualitatively agrees with our finding. In another recent work \citet{LeBrunetal2017MNRAS} make use of cosmo-OWLS simulation \citep{LeBrunetal2014MNRAS, McCarthyetal2014MNRAS} to explore the scatter and evolution of the scaling relations between the properties of the hot gas inside galaxy clusters. Their result for the evolution of the scaling relation between halo mass and bolometric luminosity agrees well with our result. \citet{RGetal2016MNRAS} investigated different observable properties of SMBH to study their evolution from an early time to the current epoch using the EAGLE simulation \citep{Schayeetal2015MNRAS, Crainetal2015MNRAS}. In this paper the authors have reported a rapid growth of black hole mass in the halo mass range $10^{11.5}-10^{12.5}M_{\odot}$. This active growth continues untill the BH reaches $10^{8}M_{\odot}$. Also in the higher mass halo BH grows linearly. We do not see such a trend mainly because we limit our sample with a BH mass well above the seed mass and also for halo mass $>10^{12}M_{\odot}$. However, they found a very little redshift evolution of this scaling relation which is consistent with our results. \citet{Barnesetal2017MNRAS} studied the global properties of galaxy clusters as well as the profiles of the hot gas inside them. They also studied X-ray and SZ properties of the hot gas inside the clusters and matched them with observations. They have used Cluster-EAGLE simulation project which is a collection of zoom simulations of formation of thirty galaxy clusters. Their density and temperature profiles showing less dense and hotter temperature of the gas at the central region again is in accordance with our claim of AGN feedback activity. 

Our study conclusively establishes the effect of AGN feedback and its implication on diffuse gas in galaxy groups and clusters and corroborates several previous other observational and theoretical work. Although the conclusion is robust we would like to stress that the conclusion might be sensitive to the particular sub-grid model of AGN feedback that is used in this simulation. Hence comparison with observational results is absolutely essential to establish the fidelity of our studies. Starting from the effect of AGN feedback on its surrounding hot gas, we study the evolution of AGN with their host halos. We elaborate new observations to verify the claims presented in our work and propose to conduct a detailed comparison between observed and simulated data in future.

\section*{Acknowledgements}

We thank the referee for some very important comments which helped in improving the draft. SC acknowledges financial support from the Department of Science and Technology through the SERB Early Career Research grant and Presidency University through the Faculty Research and Professional Development Funds. SC is grateful to the Inter University Center for Astronomy and Astrophysics (IUCAA) for providing infra-structural and financial support along with local hospitality through the IUCAA-associateship program. RKC acknowledges Department of Science and Technology for the INSPIRE fellowship. RKC is greatfulful to IUCAA, Dr. Saumyadip Samui and Dr. Ritaban Chatterjee for computational support. RKC thanks Alankar Dutta for his codes used in reading the snapfiles of K15 simulation. RKC and SC thank Andrea Lapi and members of the PRESIPACT research group for useful discussions about this work.

NK is supported by the Ramanujan Fellowship awarded by the Department of Science and Technology, Government of India. The simulations were run on the Cray XT5 supercomputer Kraken at the National Institute for Computational Sciences. The analysis was partially done on the xanadu cluster funded by the Ramanujan Fellowship at NISER. NK and TDM acknowledge support from National Science Foundation (NSF) PetaApps program, OCI-0749212 and by NSF AST-1009781.

\bibliographystyle{apj}
\bibliography{rkc}

\end{document}